\documentclass[%
reprint,
superscriptaddress,
frontmatterverbose,
showpacs,preprintnumbers,
 amsmath,amssymb,
 aps,
prx
]{revtex4-1}
\usepackage{subfigure}
\usepackage{xcolor}
\usepackage{graphicx}
\usepackage{dcolumn}
\usepackage{bm}
\usepackage{amsmath}
\usepackage{physics}
\usepackage{float}
\usepackage[utf8]{inputenc}
\usepackage[linkcolor=black,colorlinks=true,citecolor=black,filecolor=black]{hyperref}
\usepackage{academicons}
\definecolor{orcidlogocol}{HTML}{A6CE39}
\usepackage{hyperref}

\begin{document}

\newcommand{\orcid}[1]{\href{https://orcid.org/#1}{\includegraphics[width=8pt]{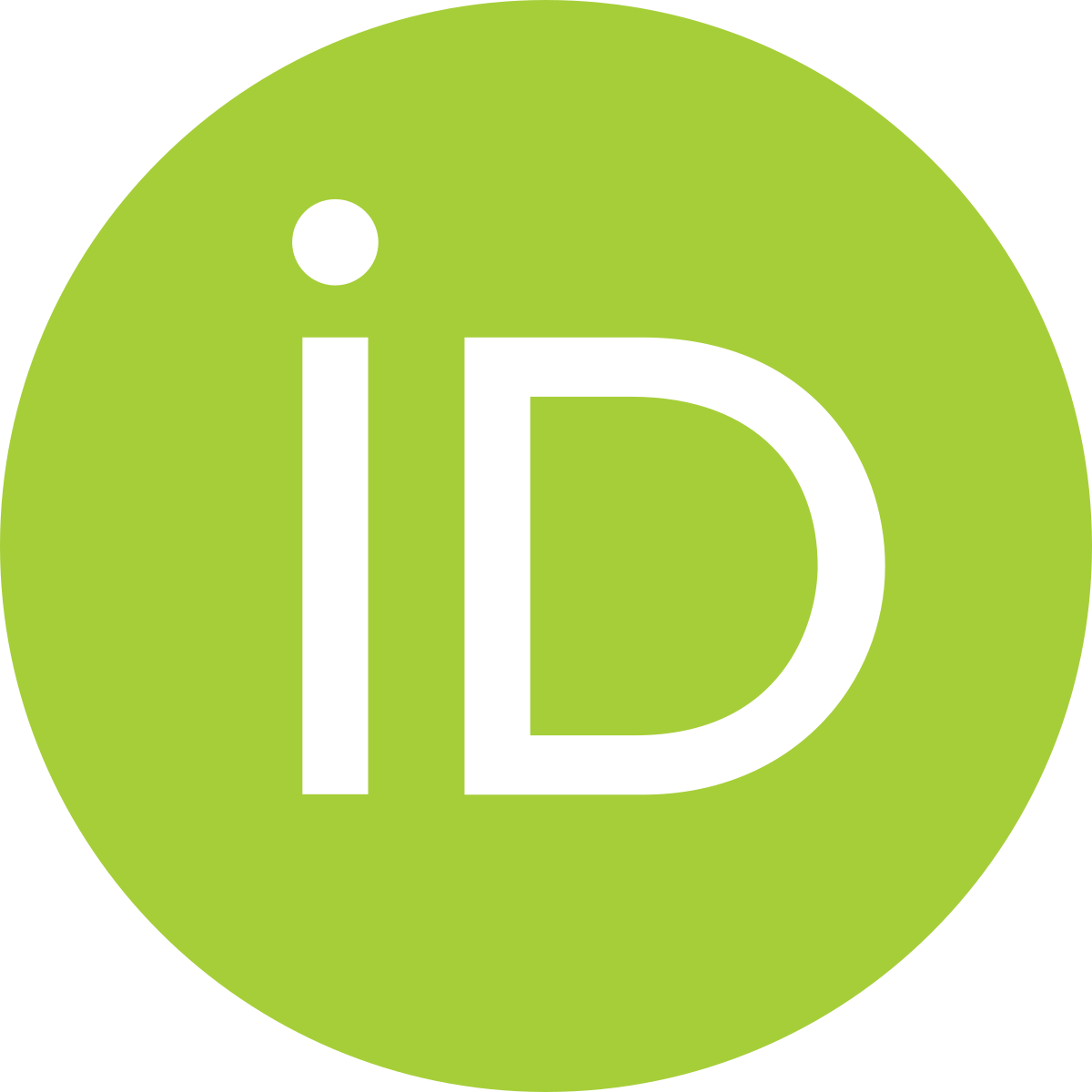}}}
\title{Generalized Entanglement Measure for Continuous Variable Systems}

\author{S Nibedita Swain \orcid{0000-0003-2326-2364}}

\email{nibedita.iiser@gmail.com}
\affiliation{Department of Physical Sciences, Indian Institute of Science Education and Research Kolkata, Mohanpur, 741246, West Bengal, India}
\author{Vineeth S. Bhaskara \orcid{0000-0002-0071-4842}}
 \email{bhaskaravineeth@gmail.com}
\affiliation{Samsung AI Centre Toronto, 101 College St Suite 420, Toronto, ON M5G 1L7, Canada}
\author{Prasanta K. Panigrahi \orcid{0000-0001-5812-0353}}%
\email{pprasanta@iiserkol.ac.in}
\affiliation{Department of Physical Sciences, Indian Institute of Science Education and Research Kolkata, Mohanpur, 741246, West Bengal, India}      
\date{\today}

\date{\today}

\begin{abstract}
Concurrence introduced by Hill and Wootters [Phys. Rev. Lett. \textbf{78}, 5022 (1997)], provides an important measure of entanglement for a general pair of qubits that is strictly positive for entangled states and vanishes for all separable states. We present an extension of entanglement measure to general pure continuous variable states of multiple degrees of freedom by generalizing the Lagrange's identity and wedge product framework proposed by Bhaskara and Panigrahi [Quantum Inf. Process. \textbf{16}, 118 (2017)] for pure discrete variable systems in arbitrary dimensions and extending the concept to mixed continuous variable states. A family of faithful entanglement measures is constructed that admit necessary and sufficient conditions for separability across arbitrary bipartitions presented by Vedral et al. [Phys. Rev. Lett. \textbf{78}, 2275 (1997)]. 
The computed entanglement measure in the present approach for general Gaussian states, pair-coherent states and non-Gaussian continuous variable Bell states, matches with known results.
We also quantify entanglement of phase randomized squeezed states and superposition of squeezed states. Our results also simplify several results in quantum entanglement theory.
\end{abstract}

\maketitle
\section{Introduction}
Quantum entanglement, having played a fundamental role in quantum information theory, is also finding its context in deeper questions, including, on the origin of space-time \cite{cowen2015quantum}, quantum field theories \cite{calabrese2004entanglement}, many-body physics \cite{amico2008entanglement}, Berry phase \cite{berry2010geometric} and quantum gravity \cite{bose2017spin}. 
Detecting the presence of such a resource and quantifying it faithfully for the general case of continuous variable systems would have far reaching applications beyond quantum computation.

Previous works, including, the extensions of Peres-Horodecki criteria by Simon \cite{simon2000peres}, Agarwal \cite{agarwal2005inseparability}, Werner \cite{werner2001bound}; non-linear maps on matrices by Giedke et al. \cite{giedke2001entanglement}; criteria based on uncertainty principles by Duan et al. \cite{duan2000inseparability}, Hillery, Nha and Zubairy \cite{hillery2006entanglement, nha2008uncertainty}, for continuous variable (CV) systems provided necessary conditions for class of non-Gaussian states for manifesting quantum optics. 
Note, however, various geometry based approaches exist to quantify entanglement \cite{bennett1996mixed, bhaskara2017generalized,guhne2021geometry,banerjee2020quantifying,banerjee2019minimum,roy2021geometric}. In this paper, we provide a family of faithful measures of entanglement, as an extension to concurrence \cite{bhaskara2017generalized}, admitting necessary and sufficient criteria for measure of entanglement \cite{vedral1997quantifying} across arbitrary bipartitions and degrees of freedom for general pure and mixed CV states, identifying an inherent geometry of entanglement in a similar spirit with examples of general Gaussian states,  phase-matched squeezed states, pair coherent states, superposition of squeezed states and non-Gaussian CV Bell states. We comment on the connections to the widely used measures in the concluding section. 
\section{PRELIMINARIES}
We introduce the notion of genuine entanglement measure based on the wedge product and the Lagrange–Brahmagupta identity  for discrete variable systems.
In an $n$-dimensional complex space, the vectors $\Vec{p}$ and $\vec{q}$ can be written as $\vec{p}=\sum_{i}p_{i}e_{i}$ and $\vec{q}=\sum_{j}q_{j}e_{j}$ respectively. The bivector $\vec{p}\wedge\vec{q}$ represents an oriented parallelopiped with sides as vectors $\vec{p}$ and $\vec{q}$:
\begin{equation*}
\vec{p}\wedge\vec{q}=\sum_{i<j}\left(p_{i}q_{j}-p_{j}q_{i}\right)e_{i}\wedge e_{j}
\end{equation*}
The Lagrange–Brahmagupta identity takes the form 
$$ \|\overrightarrow{a}\|^2 \|\overrightarrow{b}\|^2- |\overrightarrow{a} \cdot \overrightarrow{b}|^2=\|\overrightarrow{a} \wedge \overrightarrow{b} \|^2 $$ for vectors $\vec{a}$, $\vec{b}$ in $\mathbb{C}^m$.
 Without loss of generality, one can take $d_{A} = dim(\mathcal{H}_{A}) \leq dim(\mathcal{H}_{B})=d_{B}$. If \{${|\phi^{'}_{i}\rangle}$\, $i$ = $1,2,.,d_{A}$\} is an orthonormal basis of $\mathcal{H}_{A}$, it follows that
\begin{equation*}
    |\Psi^{'}\rangle = \sum_{i = 0}^{d_{\mathcal{A}}}|\phi^{'}_{i}\rangle\langle\phi^{'}_{i}|\Psi^{'}\rangle
\end{equation*}

Entanglement measure of the bi-partition $A|B$ defined in terms of wedge products \cite{bhaskara2017generalized} of post-measurement vectors can be written as 
\begin{equation*}
    C_{A|B}^{2} = 4\sum_{i<j}|\langle \phi^{'}_{i}|\Psi^{'}\rangle\wedge\langle \phi^{'}_{j}|\Psi^{'}\rangle|^{2}
\end{equation*}
where $i$ and $j$ take values from $0$ to $d_{A}$. The condition for separability across this bi-partition is $C_{A|B}^{2} = 0$.

Maximal $C_{A|B}$ for a particular bi-partition $A|B$ will correspond to the following conditions:
\begin{align*}
      \langle \phi^{'}_{i}|\Psi^{'}\rangle^{\dagger}\langle \phi^{'}_{j}|\Psi^{'}\rangle = 0 \quad \forall\quad i \neq j\\ 
        |\langle \phi^{'}_{i}|\Psi^{'}\rangle| = |\langle \phi^{'}_{j}|\Psi^{'}\rangle| \quad \forall \quad i,j
\end{align*}
\textbf{Two qubit case:}
For a two qubit system, the general state $|\psi\rangle$ in the computational basis is given by
\begin{equation*}
|\psi^{'}_{AB}\rangle=p|00\rangle+q|01\rangle+r|10\rangle+s|11\rangle
\end{equation*}
where,
$|i j\rangle=\left|i_{A}\right\rangle\otimes\left|j_{B}\right\rangle$ and $a,b,c,d\in\mathbb{C}$ satisfying the normalisation condition:
\begin{equation*}
    |p|^{2}+|q|^{2}+|r|^{2}+|s|^{2} = 1
\end{equation*}
The generalized concurrence measure in terms of wedge product (as a measure of entanglement) for $\mid\psi^{'}\rangle_{AB}$ has been obtained earlier as \cite{bhaskara2017generalized, roy2021geometric},
\begin{equation*}
\mathcal{E}=2\left|\left\langle 0_{A}\mid\psi^{'}\right\rangle \wedge\langle1_{A}|\psi^{'}\rangle\right|
\end{equation*}
These two conditions leads to the the general form of maximally entangled states for a two qubit system.
\begin{gather*}
\Bar{a}c+\Bar{b}d=0 \\
|a|^{2}+|b|^{2}=|c|^{2}+|d|^{2}
\end{gather*}
We can also get the Bell states
\begin{equation*}
 |\psi^{'}_{\pm}\rangle = \frac{(\left|00\right\rangle \pm  \left|11\right\rangle)}{\sqrt{2}}
\end{equation*}
\begin{equation*}
 |\phi^{'}_{\pm}\rangle = \frac{(\left|01\right\rangle \pm  \left|10\right\rangle)}{\sqrt{2}}
\end{equation*}

\textbf{The three qubit case:}
The  general state is given by 
\begin{equation*}
    \begin{aligned}
    |\psi^{'}_{ABC}\rangle = a_{0}|000\rangle + a_{1}|001\rangle +  a_{2}|010\rangle + a_{3}|011\rangle + a_{4}|100\rangle\\ + a_{5}|101\rangle + a_{6}|110\rangle + a_{7}|111\rangle
    \end{aligned}
\end{equation*}
where, $a_{i}\in C$, $i = 1-7$  satisfy the normalisation condition $\sum_{i = 0}^{7}|a_{i}|^{2}=1$.
The measure of entanglement $\mathcal{E}$ is given by sum of concurrence corresponding to all three bipartitions    \cite{bhaskara2017generalized}
\begin{equation*}
    \mathcal{E} = \mathcal{E}_{A|BC} + \mathcal{E}_{B|AC} + \mathcal{E}_{C|AB}
\end{equation*}
In the wedge product formalism, we can write the concurrence as:
\begin{equation*}
    \mathcal{E} = 2\sum_{i = A,B,C}|\langle 0_{i}|\psi^{'}_{ABC}\rangle \wedge \langle 1_{i}|\psi^{'}_{ABC}\rangle|
\end{equation*}
$|\Psi^{'}_{ABC}\rangle$ can be written in the following form:
\begin{equation*}
    |\Psi^{'}_{ABC}\rangle = |0_{A}\rangle\langle 0_{A}|\Psi^{'}_{ABC}\rangle +|1_{A}\rangle\langle 1_{A}|\Psi^{'}_{ABC}\rangle
\end{equation*}
The conditions for maximally entangled states are : $\langle 0_{i}| \psi^{'}_{ABC}\rangle$ must be orthogonal to $\langle 1_{i}|\psi^{'}_{ABC}\rangle$  and $|\langle 0_{i}|\psi^{'}_{ABC}\rangle| = |\langle 1_{i}|\psi^{'}_{ABC}\rangle|$ for each $i = A, B, C$ 
This maximally entangled state (GHZ state) by using the above conditions is defined as,
  \begin{equation*}
    |\psi^{'}\rangle=\frac{1}{\sqrt{2}}(|000\rangle+|111\rangle)
\end{equation*}
We then explicitly explain  in the next section that for continuous variable systems a much more refined and efficient form of generalized entanglement measure (GEM) based on extended wedge product and the Lagrange–Brahmagupta identity. The main advantage of our proposed measure GEM consists
in a reduced computational effort required for its evaluation.
\section{Generalized Entanglement Measure (GEM) for pure CV states}
We define separability for pure states in the context of CV systems for future convenience. Consider a $n$-degree of freedom quantum system. Let $P|Q$ be a bipartition across the degrees of freedom of this composite(whole) system $P\cup Q$, with respective infinite-dimensional Hilbert spaces $\mathcal{H}_P$ and $\mathcal{H}_Q$ for the states of the sub-systems $P$ and $Q$, then the state space of the composite system is given by the tensor product $\mathcal{H}=\mathcal{H}_P \otimes \mathcal{H}_Q$. If a pure state $|\psi\rangle \in \mathcal{H}$ of the composite system with $\rho_\psi= |\psi \rangle \langle \psi |$ can be written in the form 
\[|\psi\rangle=|\phi\rangle \otimes |\chi\rangle,\text{ i.e., } \rho_\psi = \rho_\phi \otimes \rho_\chi, \] where $|\phi\rangle \in \mathcal{H}_P$ and $|\chi\rangle \in \mathcal{H}_Q$ are the pure states of the sub-systems $P$ and $Q$ respectively with $\rho_\phi= |\phi \rangle \langle \phi |$ and $\rho_\chi= |\chi \rangle \langle \chi |$, then the system is said to be separable across the bipartition $P|Q$. Otherwise, the sub-systems $P$ and $Q$ are said to be entangled.

Consider a general $n$-degree of freedom pure CV state $|\psi\rangle$ with the degrees of freedom taking continuous values and labeled by \{$x_1,~x_2,~\ldots,~x_n$\} in an orthonormal basis with $\langle \vec{x'}|\vec{x}\rangle = \delta(\vec{x}-\vec{x'})$ and $\int |\vec{x}\rangle \langle \vec{x}|~ \mathrm{d}\vec{x}=1$ as
\begin{equation}
|\psi\rangle = \int \phi(x_1,...,x_n) ~~|x_1\rangle \otimes |x_2\rangle \otimes \cdots \otimes |x_n\rangle ~~ \mathrm{d}\vec{x}, \label{fullcvstate}
\end{equation}
with $\langle \psi|\psi\rangle = 1$, i.e.,
\[\int {\phi}^*(x_1,...,x_n) \phi(x_1,...,x_n) ~\mathrm{d}\vec{x} = 1,\]
 where $\vec{x}=(x_1,x_2,...,x_n)$, $\mathrm{d}\vec{x} \equiv \mathrm{d}^nx = \mathrm{d}x_1 ~ \mathrm{d}x_2~ ...~ \mathrm{d}x_n$, and $\delta$ is the Dirac delta function of appropriate dimension. Note that the limits of the integrals are over the appropriate continuous range of values for the degrees of freedom (commonly, $-\infty$ to $+\infty$) unless otherwise specified. By $n$-degree of freedom system one could mean, for instance, a system of $n$-particles in one spatial dimension, or a system of $k$-particles in $3D$ where $n=3k$, or a quantum optics system having multiple modes. The physical state $|\psi\rangle$ exists in an infinite-dimensional Hilbert space spanned by $\{|x\rangle\}$. Note that, unlike the case of discrete variable (DV) systems, the basis states $\{|x\rangle\}$ by themselves are not normalizable and hence non-physical. 
 
The generalized entanglement measure (GEM) for pure CV states will now be defined as
\begin{widetext}
    \begin{equation}
    \begin{aligned}
      \mathcal{E}_{\mathcal{M}}^{2} = 2\Big[1 - \iint \Big|\int \phi(y_{1}^{'},y_{2}^{'},..,y_{m}^{'},x_{m+1},..,x_{n})\phi^{*}(y_{1},y_{2},..,y_{m},x_{m+1},..,x_{n}) d^{n-m}x   \Big|^{2}  d^{m}y d^{m}y^{'} \Big].
    \end{aligned}
\end{equation}
\end{widetext}

We can define,
\begin{equation}
 \Tilde{\Phi}(\mathcal{X}) = \braket{\mathcal{X}|\mathbb{N}_{1}^{T}}{\psi} \braket{\mathcal{X}|\mathbb{N}_{2}^{T}}{\psi} = \phi(\mathbb{N}_{1}\mathcal{X})\phi (\mathbb{N}_{2}\mathcal{X})   
\end{equation}
$\mathbb{N}_{1} = \begin{bmatrix}
 1_{n\times n}  & 0_{n\times n}
\end{bmatrix}_{n\times 2n}$ , $\mathbb{N}_{1} = \begin{bmatrix}
 0_{n\times n}  & 1_{n\times n}
\end{bmatrix}_{n\times 2n}$, 
\begin{equation*}
  \mathbb{M} = \begin{bmatrix}
             1_{m\times m} &  \\
               &  0_{(n-m)\times (n-m)}
\end{bmatrix}_{2n\times 2n}  
\end{equation*}
\begin{equation*}
  \wedge_{m} = \begin{bmatrix}
            1_{n \times{n}}-\mathbb{M}_{n\times n} & \mathbb{M}_{n\times n}\\
            \mathbb{M}_{n\times n}  & 1_{n\times n} - \mathbb{M}_{n\times n}
\end{bmatrix}_{2n\times 2n}  
\end{equation*}

Eq. (5) can be written in terms of $\phi$ and $\wedge_{m}$ as

\begin{equation}
   \mathcal{E}_{\mathcal{M}}^{2} = 2\Big[ 1- \textrm{Re}\int  \Tilde{\Phi}(\mathcal{X})\Tilde{\Phi}^{*}(\mathcal{\wedge}_{m}\mathcal{X})\textrm{d}^{2n}\mathcal{X}\Big]
\end{equation}

\textbf{Preposition}: The state $\psi$ is said to be separable across the bipartition $\mathcal{M}|\overline{\mathcal{M}}$ if and only if $\psi$ is expressible as
\begin{equation}
   \ket{\psi} = \int \phi(x_{1},x_{2},....,x_{n})\ket{x_{1}}\otimes\ket{x_{2}}\otimes...\otimes\ket{x_{n}} d\Vec{x}
\end{equation}
\begin{align*}
=\bigg[\int \phi_{\mathcal{M}}&(x_1,...,x_m)~|x_1...x_m\rangle~\mathrm{d}^mx \bigg] ~\otimes~ \nonumber \\ 
&\bigg[\int \phi_{\overline{\mathcal{M}}}(x_{m+1},...,x_n)~|x_{m+1}...x_n\rangle~\mathrm{d}^{n-m}x \bigg],
\end{align*}
Proof: 
Consider the bipartite separability of a particular set $\mathcal{M}$ of $m$ degrees $(m<n)$ out of the $n$ degrees of freedom of the system. Without any loss of generality, let the $m$-degrees be labeled by \{$1,2,...,m$\}, so that the degrees labeled by \{$m+1,m+2,...,n$\} represent the rest of $(n-m)$-degrees of freedom belonging to the complement set $\overline{\mathcal{M}}$.
The state $|\psi\rangle$ is said to be separable across the bipartition $\mathcal{M}|\overline{\mathcal{M}}$ if and only if $|\psi\rangle$ is expressible as
\begin{align*}
\bigg[\int \phi_{\mathcal{M}}&(x_1,...,x_m)~|x_1...x_m\rangle~\mathrm{d}^mx \bigg] ~\otimes~ \nonumber \\ 
&\bigg[\int \phi_{\overline{\mathcal{M}}}(x_{m+1},...,x_n)~|x_{m+1}...x_n\rangle~\mathrm{d}^{n-m}x \bigg],
\end{align*}
where $\phi_{\mathcal{M}}$ is the normalized pure state of the sub-system $\mathcal{M}$, $|x_1...x_m\rangle \equiv |x_1\rangle \otimes \cdots \otimes |x_m\rangle$, $\mathrm{d}^mx \equiv \mathrm{d}x_1~\mathrm{d}x_2~...~\mathrm{d}x_m$, and similarly $\phi_{\overline{\mathcal{M}}}$ is the normalized pure state of the sub-system $\overline{\mathcal{M}}$, $|x_{m+1}...x_n\rangle \equiv |x_{m+1}\rangle \otimes \cdots \otimes |x_n\rangle$, $\mathrm{d}^{n-m}x \equiv \mathrm{d}x_{m+1}~\mathrm{d}x_{m+2}~...~\mathrm{d}x_n$.

One may rewrite the state $|\psi\rangle$, defined in Eq. \eqref{fullcvstate}, as
\begin{equation*}
\int \bigg[ |x_1...x_m\rangle~ \otimes ~ \bigg( \int \phi(x_1,...,x_n) ~|x_{m+1}...x_n\rangle~\mathrm{d}^{n-m}x \bigg)~\mathrm{d}^mx \bigg] \label{eq2}.
\end{equation*}
By noting that 
\begin{align*}
&\langle x'_1x'_2...x'_m|\psi\rangle \nonumber\\
&=\int \phi(x_1,...,x_n)~ \underbrace{\langle x'_1x'_2...x'_m|x_1x_2...x_n\rangle}_{\delta(x_1-x'_1,...,x_m-x'_m)~|x_{m+1}...x_n\rangle} ~\mathrm{d}^nx \\
&=\int \phi(x'_1,x'_2,...,x'_m,x_{m+1},...,x_n)~|x_{m+1}...x_n\rangle  ~\mathrm{d}^{n-m}x, \label{bradpsi1}
\end{align*}
one may express $|\psi\rangle$ as 
\begin{equation}
|\psi\rangle = \int \bigg[ |x_1...x_m\rangle~ \otimes ~ \bigg( \langle x_1x_2...x_m|\psi\rangle \bigg)~\mathrm{d}^mx \bigg]. \label{eqMAIN1}
\end{equation}

Observe that, for the separability of $|\psi\rangle$ across $\mathcal{M}|\overline{\mathcal{M}}$, each of the vectors $\langle x_1x_2...x_m|\psi\rangle$ in Eq. \eqref{eqMAIN1} must be mutually ``parallel" for the $m$-degree of freedom state to factor out, i.e., for each $\vec{r}=(r_1,r_2,...,r_m)$ one needs
\[ \langle r_1r_2...r_m|\psi\rangle = c_{\vec{r}\vec{s}} \langle s_1s_2...s_m|\psi\rangle \]
for any $\vec{s}=(s_1,s_2,...,s_m)$ where $c_{\vec{r}\vec{s}}$ is some complex scalar, for separability. This becomes evident once one chooses, say, $(s_1,s_2,...,s_m)=(0,0,...,0)=\vec{0}$, so that 

\[\langle r_1r_2...r_m|\psi\rangle = c(r_1,r_2,...,r_m) ~\langle \underbrace{00...0}_{m}|\psi\rangle \]
where $c(r_1,r_2,...,r_m)$ is some complex scalar. Substituting this back in Eq. \eqref{eqMAIN1}, one can see the state becomes separable (with constant $k$ ensuring the normalization of each of the sub-system's state) as
\begin{align*}
|\psi\rangle &= \int \bigg[ |x_1...x_m\rangle~ \otimes ~c(x_1,x_2,...,x_m) ~\langle 00...0|\psi\rangle ~\mathrm{d}^mx \bigg] \nonumber \\
&= \bigg(\underbrace{k\int|x_1...x_m\rangle~c(x_1,x_2,...,x_m)  ~\mathrm{d}^mx}_{state~ of~ \mathcal{M}} \bigg)\otimes\underbrace{\frac{1}{k}\langle 00...0|\psi\rangle,}_{state~ of~ \overline{\mathcal{M}}} \label{decomposeSection}
\end{align*}
\begin{widetext}
\begin{equation}
  \ket{\psi} = \Big( \int \phi(x_{1},...,x_{m})\ket{x_{1}...x_{m}} d^{m}x\Big) \otimes \Big(\phi(0,..0,x_{m+1},..,x_{n})\ket{x_{m+1}..x_{n}} d^{n-m}x          \Big) 
\end{equation}
\end{widetext}

Interestingly, therefore, one may note that even if a single ``pair" of elements of the continuous, infinite set of vectors \{$\langle x_1x_2...x_m|\psi\rangle$\} over the continuous variables $x_1,...,x_m$ is not mutually parallel, this adds to the presence of entanglement. 
We express this condition for separability using the notion of a wedge product extended to multivariable complex-valued function spaces based on the framework proposed in Ref. \cite{bhaskara2017generalized} for general pure discrete variable systems in arbitrary dimensions.

\textbf{Theorem}: If a single pair of elements of the continuous, infinite set of vectors ${\braket{x_{1}...x_{m}}{\psi}}$ over the continuous variables $x_{1},..x_{m}$ is not mutually parallel, the family of faithful measures of entanglement  across the bipartition $\mathcal{M}|\overline{\mathcal{M}}$ is defined as
\begin{widetext}
\begin{equation}
   \mathcal{E}_{\mathcal{M}}^{2} = 2\Bigg[1 -  \iint \Bigg | \int \phi (y_{1}^{'},y_{2}^{'},..,y_{m}^{'},x_{m+1},..x_{n}) \phi^{*} (y_{1},y_{2},..,y_{m},x_{m+1},..x_{n})d^{n-m}x \Bigg|^{2} d^{m}y d^{m}y^{'}             \Bigg] 
\end{equation}
\end{widetext}
Proof: 

In geometric algebra \cite{doran2003geometric}, the wedge product of two vectors is seen as a particular generalization of cross product to higher dimensions. We construct such a notion for the case of complex infinite-dimensional vector spaces. Consider two vectors $\vec{a}$ and $\vec{b}$ in the complex, infinite-dimensional space as
\[\vec{a}=\int f(\vec{x}) ~|\vec{x}\rangle ~\mathrm{d}\vec{x}, ~~\vec{b}=\int g(\vec{x}) ~|\vec{x}\rangle~ \mathrm{d}\vec{x}  \]
in the continuous orthonormal basis set $\{|\vec{x}\rangle\}$ with $\langle \vec{x'}|\vec{x}\rangle = \delta(\vec{x}-\vec{x'})$ and $\int |\vec{x}\rangle \langle \vec{x}|~ \mathrm{d}\vec{x}=1$ where $\vec{x}=(x_1,...,x_n)$. Then the wedge product of $\vec{a}$ and $\vec{b}$ in the interval $(\vec{t},\vec{u})$ is defined as a bivector in an ``exterior" space with continuous basis set $\{|\vec{x}\rangle \wedge |\vec{x'}\rangle \}_{x'>x}$, stipulating that $|\vec{x}\rangle \wedge |\vec{x'}\rangle = - |\vec{x'}\rangle \wedge |\vec{x}\rangle$ and $|\vec{x}\rangle \wedge |\vec{x}\rangle=0$, as
\begin{equation*}
\vec{a} \wedge \vec{b} = \int\limits_{\vec{x}=\vec{t}}^{\vec{u}} ~\int\limits_{\vec{x'}=\vec{x}}^{\vec{u}} \left[f(\vec{x})g(\vec{x'}) - f(\vec{x'})g(\vec{x}) \right] |\vec{x}\rangle \wedge |\vec{x'}\rangle~\mathrm{d}\vec{x'} ~\mathrm{d}\vec{x},
\end{equation*}
where $\vec{t}=(t_1,t_2,...,t_n)$, and $\vec{u}=(u_1,u_2,...,u_n)$. Therefore, one may note $\vec{a} \wedge \vec{b}=0 \iff \vec{b} = k\vec{a}$, and $\vec{a} \wedge \vec{b}=-\vec{b} \wedge \vec{a}$, by definition, for some complex scalar $k$ and vectors $\vec{a},~\vec{b}$.

This notion of an extended wedge product allows one to write the separability condition in a compact and useful form. Since one requires that each of the vectors in the continuous set \{$\langle x_1x_2...x_m|\psi\rangle$\} to be mutually ``parallel" for the separability across $\mathcal{M}|\overline{\mathcal{M}}$, their mutual wedge products must vanish, equivalently, for separability. This is a necessary and sufficient condition for separability as noted before. Hence, one may construct a family of faithful measures of entanglement, parametrized by $f$, $p$, and $q$, across the bipartition as 

\begin{multline*}
\mathcal{E}_{\mathcal{M}} = \\ \left[\int \int   f\left(\left|\left|	\langle y'_1y'_2...y'_m|\psi\rangle \wedge \langle y_1y_2...y_m|\psi\rangle \right|\right|_\text{p}\right) \mathrm{d}^my~ \mathrm{d}^my'\right]^{1/q} 
\end{multline*}
where $f:\mathbb{R}\to\mathbb{R}$ with $f(x)=0$ iff $x=0$ so that \hbox{$\mathcal{E}_{\mathcal{M}}=0$ $\iff$ separability}, in addition to $f$ being a monotonic and strictly increasing function in $\mathbb{R}^+$ and $q\in\mathbb{R}^+$ so that $\mathcal{E}_{\mathcal{M}}>0$ measures entanglement faithfully; the $p$-norm is computed in the basis $\{|x_{m+1}...x_n\rangle \wedge |x'_{m+1}...x'_n\rangle\}_{x'>x}$, and $ \langle y'_1y'_2...y'_m|\psi\rangle \wedge \langle y_1y_2...y_m|\psi\rangle \equiv$
\begin{widetext}	 
\begin{align}
\int\limits_{\vec{x}=-\infty}^{+\infty} ~\int\limits_{\vec{x'}=\vec{x}}^{+\infty} \bigg[\phi(y'_1&,y'_2,...,y'_m,x_{m+1},...,x_n)~\phi(y_1,y_2,...,y_m,x'_{m+1},...,x'_n) ~~- \nonumber \\&  \phi(y'_1,y'_2,...,y'_m,x'_{m+1},...,x'_n)~\phi(y_1,y_2,...,y_m,x_{m+1},...,x_n) \bigg]~|x_{m+1}...x_n\rangle \wedge |x'_{m+1}...x'_n\rangle ~\mathrm{d}^{n-m}x'~ \mathrm{d}^{n-m}x. \label{vineq1}
\end{align}

The Lagrange's identity takes the form

$$||\vec{a}||^{2}||\vec{b}||^{2} - |\vec{a}.\vec{b}|^{2} = ||\vec{a}\wedge \vec{b}||^{2}$$
By this identity, one may rewrite the entanglement measure $E^2_{\mathcal{M}}$ constructed as,
\begin{equation}
    \begin{aligned}
    \Big(\int_{t}^{u} |f(\vec{x})|^{2} d\vec{x}\Big) \Big(\int_{t}^{u} |g(\vec{x})|^{2} d\vec{x}\Big) - \Big |\int_{t}^{u} f(\vec{x}) g^{*}(\vec{x}) \Big|^{2} = \iint_{t}^{u} \Big|f(\vec{x}) g(\vec{x^{'}}) - f(\vec{x^{'}}) g(\vec{x})\Big|^{2} d\vec{x^{'}} d\vec{x} \nonumber
    \end{aligned}
\end{equation}

\begin{equation}
    \begin{aligned}
      \mathcal{E}_{\mathcal{M}}^{2} = 2\iint \Big[\Big( \int |\phi(y_{1}^{'}y_{2}^{'}..y_{m}^{'},x_{m+1},..,x_{n})|^{2} d^{n-m}x\Big)  \Big( \int |\phi(y_{1}y_{2}..y_{m},x_{m+1},..,x_{n})|^{2} d^{n-m}x\Big)  - \\ \Big| \int \phi(y_{1}^{'}y_{2}^{'}..y_{m}^{'},x_{m+1},..,x_{n})\phi^{*}(y_{1}y_{2}..y_{m},x_{m+1},..,x_{n}) d^{n-m}x \Big|^{2}\Big] d^{m}y d^{m}y^{'} \nonumber
    \end{aligned}
\end{equation}

\begin{equation}
    \begin{aligned}
      \mathcal{E}_{\mathcal{M}}^{2} = 2\Big[1 - \iint \Big|\int \phi(y_{1}^{'}y_{2}^{'}..y_{m}^{'},x_{m+1},..,x_{n})\phi^{*}(y_{1}y_{2}..y_{m},x_{m+1},..,x_{n}) d^{n-m}x   \Big|^{2}  d^{m}y d^{m}y^{'}\Big]
    \end{aligned}
\end{equation}
\end{widetext}
noting the normalization of $\phi$. This may elegantly be written in terms of $\Phi$ and $\Lambda_m$ as 
\begin{align}
\mathcal{E}^2_{\mathcal{M}}= 2 \left[  1 - \text{Re}\int \Phi(\text{X})~ \Phi^*(\Lambda_m \text{X}) ~ \mathrm{d}^{2n}\text{X} \right].
\end{align}
Hence, for maximal entanglement, one needs $\Phi(\text{X})$ and $\Phi(\Lambda_m \text{X})$ to be orthogonal, i.e., their inner product must vanish, so that $\mathcal{E}^2_{\mathcal{M}}$ takes the maximum value of $2$. On the contrary, when $\Phi(\text{X})=\Phi(\Lambda_m \text{X})$, their inner product takes the maximum overlap of $1$, thereby, implying separability with $\mathcal{E}^2_{\mathcal{M}}=0$. This is one of the important results of the paper on the geometry of entanglement in CV pure systems.

\subsection{Gaussian CV states}
We consider the example of a general pure Gaussian CV state to evaluate our criterion and provide the condition for separability, and analyze GEM for the case of a general two-mode Gaussian states.
\begin{equation}
\psi_{1}(x_{1},..,x_{n}) = \mathcal{N}_{1} exp\Big(-\frac{1}{2}[\sum_{k=1}^{n}a_{k}x_{k}^{2} + \sum_{k,j;j>k}^{n} c_{kj}x_{k}x_{j}]\Big)
\end{equation} 
Where $a_{k} \in \mathbb{R}, \: a_{k} \ge 0, \: c_{kj} \in \mathbb{C},$ and $\mathcal{N}_{1}$ is the appropriate normalization term for the wavefunction. Consider the separability of m degrees labeled by ${x_{1},..,x_{m}}$ from the n available degrees of freedom. For separability across the $m|(n-m)$ bipartition, one needs
\begin{multline*}
    \exp\Big(-\frac{1}{2}[\sum_{k=1}^{m}a_{k}y^{'2}_{k} + \sum_{k=m+1}^{n} a_{k}x_{k}^{2} + \sum_{k,j=1;j>k}^{k,j=m} c_{kj}y^{'}_{k}y^{'}_{j} + \\ \sum_{k=1,j=m+1;j>k}^{k=m,j=n} c_{kj}y^{'}_{k}x_{j} + \sum_{k,j=m+1;j>k}^{k,j=n} c_{kj}x_{i}x_{j}  ]\Big) \cross
\end{multline*}
\begin{multline*}
    \exp\Big(-\frac{1}{2}[\sum_{k=1}^{m}a_{k}y^{2}_{k} + \sum_{k=m+1}^{n} a_{k}x_{k}^{'2} + \sum_{k,j=1;j>k}^{k,j=m} c_{kj}y_{k}y_{j} + \\ \sum_{k=1,j=m+1}^{k=m,j=n} c_{kj}y_{k}x_{j}^{'} + \sum_{k,j=m+1;j>k}^{k,j=n} c_{kj}x^{'}_{i}x^{'}_{j}  ]\Big)
\end{multline*}
\begin{multline*}
    =\exp\Big(-\frac{1}{2}[\sum_{k=1}^{m}a_{k}y^{'2}_{k} + \sum_{k=m+1}^{n} a_{k}x_{k}^{'2} + \sum_{k,j=1;j>k}^{k,j=m} c_{kj}y^{'}_{k}y^{'}_{j} + \\ \sum_{k=1,j=m+1}^{k=m,j=n} c_{kj}y^{'}_{k}x^{'}_{j} + \sum_{k,j=m+1;j>k}^{k,j=n} c_{kj}x^{'}_{i}x^{'}_{j}]\Big) \cross
\end{multline*}
\begin{multline*}
    \exp\Big(-\frac{1}{2}[\sum_{k=1}^{m}a_{k}y^{2}_{k} + \sum_{k=m+1}^{n} a_{k}x_{k}^{2} + \sum_{k,j=1;j>k}^{k,j=m} c_{kj}y_{k}y_{j} + \\ \sum_{k=1,j=m+1}^{k=m,j=n} c_{kj}y_{k}x_{j} + \sum_{k,j=m+1;j>k}^{k,j=n} c_{kj}x_{i}x_{j}  ]\Big) 
\end{multline*}
This simplifies to the requirement
\begin{multline*}
  \sum_{k=1,j=m+1}^{k=m,j=n} c_{k j}(y^{'}_{k}x_{j}+y_{k}x_{j}^{'}) \\ =\sum_{k=1,j=m+1}^{k=m,j=n} c_{k j}(y^{'}_{k}x^{'}_{j}+y_{k}x_{j})   
\end{multline*}
 that can only be true for arbitrary values of $ y_{i}^{'},\: x_{i},\: x_{i}^{'}, y_{i},$ iff
\begin{align*}
   c_{kj} & = 0 \:\forall \: k \in [1,m]\: \textrm{and} \: j\in [m+1, n], \textrm{or}\\
   \mathbb{V}_{k,j} & =  \mathbb{V}_{k,j}^{T} = 0 \: \forall \: k \in [1,m] \:\textrm{and}\: j\in [m+1, n], \textrm{or} 
  \end{align*}
\begin{equation}
     \mathbb{M}\mathbb{V}(1_{n\cross n} - \mathbb{M})  = \mathbb{M}\mathbb{V}^{T} (1_{n\cross n} - \mathbb{M}) = 0_{n \cross n}
\end{equation}
Where $\mathbb{V} = \sum^{-1}$ is the inverse of the covariance matrix of the Gaussian. This is a necessary and sufficient condition for separability of the general Gaussian wavefunction $\psi_{1}(\vec{x})$ in n-degrees of freedom across the bipartitions $m|(n-m)$. Moreover, one may say that the system is entangled across the bipartitions iff $\exists \: k,j \: \textrm{such that} \: c_{kj} \neq 0$ for some $k \in [1,m] \: \textrm{and} \: j \in [m+1,n].$
 
 To analyze GEM for the case general two-mode Gaussian state,
 \begin{equation}
     \psi_{2}(x_{1},x_{2}) = \mathcal{N}_{2}e^{-\frac{1}{2}(ax_{1}^{2} + bx_{2}^{2} + cx_{1}x_{2})}
 \end{equation}
where $a,b \in \mathbb{R}, a,b >0,$ $c$ is either purely real or imaginary, and
\begin{align*}
    \mathcal{N}_{2} & = [\frac{2\pi}{\sqrt{4ab-c^{2}}}]^{\frac{-1}{2}} \quad \textrm{if c is real} \\
    & = \Big[\frac{\pi}{\sqrt{ab}}\Big]^{\frac{-1}{2}} \quad \qquad \qquad \textrm{if} \: c = im,, m \in \mathbb{R}
\end{align*}
with $i \equiv \sqrt{-1}$. Using Eq. (12), one may compute the GEM across the modes as
\begin{align*}
   \mathcal{E}_{\mathcal{M}}^{2} & = 2[1 - \frac{\sqrt{4ab-c^{2}}}{2\sqrt{ab}}] \quad \: \textrm{if c is real} \\
    & = 2[1 - \frac{2\sqrt{ab}}{\sqrt{4ab + m^{2}}}] \quad  \textrm{if} \: c = im,, m \in \mathbb{R}
\end{align*}
Clearly, $E =0\: \textrm{iff} \:c =0$ for both the cases. The entanglement depends on the parameter c. When c is real, one requires $-2\sqrt{ab} < c < 2\sqrt{ab} $. So that the state remains normalizable and hence physical. For this case, as $c \to  \pm 2\sqrt{ab}, \: E^{2} \to 2$. When c is purely imaginary with $c = im, \: m\in \mathbb{R},\: \textrm{as} \: m \to \pm \infty, \: E^{2} \to 2$ asymptotically.

\subsection{The pair coherent state}
The pair coherent state \cite{agarwal2005quantitative} is given by
\begin{equation}
    \ket{\xi, 0} = \frac{1}{\sqrt{I_{0}(2|\xi|)}} \sum \frac{\xi^{i}}{i!} \ket{i,i}
  \end{equation}
where $I_{0}(2|\xi|)$ is the modified Bessel function of order zero. By using GEM for pure states, for separability across the $m|(n-m)$ bipartition, one needs
\begin{figure}[ht]
\begin{minipage}[b]{0.45\linewidth}
\centering
\includegraphics[width=\textwidth]{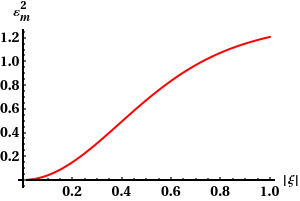}
\caption{Variation of the entanglement measure $(\mathcal{E}^{2}_{m})$ for pair coherent state with $|\xi|$, All the axes are dimensionless.}
\label{fig:figure1}
\end{minipage}
\hspace{0.45cm}
\begin{minipage}[b]{0.45\linewidth}
\centering
\includegraphics[width=\textwidth]{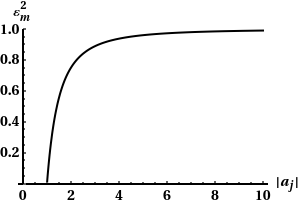}
\caption{Variation of the entanglement measure $(\mathcal{E}^{2}_{m})$ for generalized superposition state with $|a_{j}|$, All the axes are dimensionless.}
\label{fig:figure2}
\end{minipage}
\end{figure}
\begin{equation}
    \sum N_{ii} N_{jj}^{'} \ket{i,j}\bra{i,j} = \sum N_{j,i}N^{'}_{j,i} \ket{j,i}\bra{j,i}
\end{equation}
For $i \neq j$, by using Eq. (12), the entanglement measure is given by
\begin{equation}
    \mathcal{E}^{2}_{\mathcal{M}} = 2\Big[1 - \frac{1}{I_{0}(2|\xi|)^{2}} \sum \frac{|\xi|^{4i}}{i!^{4}}\Big]
\end{equation}
We plot the GEM for pair-coherent state (15) with $|\xi|$ in Fig. 1. Clearly, for non-zero values of $|\xi|$, $(\mathcal{E}^{2}_{m})$ is non-zero. This implies entanglement in pair coherent state. For small values of $|\xi|$,  $(\mathcal{E}^{2}_{m})$  increases slowly, then it saturates at larger values. 
\subsection{Superposition of squeezed states}
 In Ref. \cite{kannan2021positive}, the first kind superposition of \textit{j} squeezed vacuum state which have the same squeezing value in Fock basis is given by
\begin{equation}
    \ket{\xi}_{j} = N_{j}\sum (-e^{i\theta}\tanh{r})^{jn} \frac{\sqrt{(2jn)!}}{2^{jn}(jn)!}\ket{2jn}
\end{equation} 
where $$N_{j} = \Big(\sum \frac{(2jn)!}{2^{2jn}(jn)!}(\tanh{r})^{2jn}\Big)^{-\frac{1}{2}}$$ 
By using Eq. (12), the entanglement measure is given by
 \begin{equation}
     \mathcal{E}^{2}_{\mathcal{M}} = 0
 \end{equation}
The generalized, 2nd kind of superposition, state with different squeezing value and weight factor is 
\begin{equation}
 \ket{\psi} = \Big(\sum_{p=1}^{l}\sum_{j=1}^{l} \frac{a_{p}a_{j}^{*}}{\sqrt{\cosh{(q_{p}-q_{j})}}}\Big)^{-\frac{1}{2}} \sum_{j=0}^{l-1} a_{j}\ket{\xi_{j}}  
\end{equation}
where $\xi_{j} = q_{j}e^{i\theta_{j}}$, $a_{j}'s$ are the weight factors.
By using GEM , the entanglement measure is
\begin{equation}
  \mathcal{E}^{2}_{\mathcal{M}} = 2\Big[1- \sum\frac{1}{|a_{j}|^{2}}\Big]  
\end{equation}
For superposition of first kind, the entanglement measure is found to be 0 and from FIG. 2, the entanglement measure for generalized 2nd kind superposition state increases slowly, then saturates when weight factor reaches maximum.
\subsection{Non-Gaussian CV Bell  state}
From Ref. \cite{agarwal2005inseparability}, the non-Gaussian continuous variable state is expressed as,
\begin{equation}
    \psi_{ng} = \sqrt{\frac{2}{\pi}}(px_{1} + qx_{2})e^{-\frac{(x_{1}^{2}+x_{2}^{2})}{2}}
\end{equation}
The state is a composite system of bosonic partice formed from the ground and excited state of the harmonic oscillators \cite{agarwal1997vortex}. The experimental scheme of this state has already been proposed \cite{garcia2004proposal}. The Peres-Horodecki criterion \cite{peres1996separability, horodecki2001separability} is only sufficient for Eq. (22). Agarwal et. al. shows inseparability of the state (22) via inequalities \cite{agarwal2005inseparability} which is also applicable for the state (22). 

By using GEM, across the $m|(n-m)$ bipartitions, one finds
\begin{align*}
   N_{m}N_{m}^{*}\ket{m,n}\bra{m,n}+N_{n}N_{n}^{*}\ket{n,m}\bra{n,m}\\+N_{m}^{*}N_{n}\ket{m,m}\bra{n,n}+N_{m}N_{n}^{*}\ket{n,n}\bra{m,m}  
\end{align*}
For $m \ne n$, the measure of entanglement is given by
\begin{equation}
\mathcal{E}^{2}_{\mathcal{M}} = 2\Big[1 - \frac{2}{\pi}p^{2}q^{2}\Big] 
\end{equation}
Clearly, we can see that the state is entangled and $\mathcal{E}^{2}_{\mathcal{M}}= 0$ iff p or q is zero. The entanglement depends on the parameter p or q.

\section{GEM for mixed CV states}
We can now define the generalized entanglement measure (GEM) for a general mixed continuous variable systems. For N-degree of freedom continuous variable mixed state, the GEM is defined as,
\begin{equation}
    \mathcal{G}(\rho) = \underset{\{a_{i}, \ket{\psi^{i}}\}}{\mathrm{min}} \sum_{i} a_{i}	E^2_{\mathcal{M}}(\ket{\psi^{i}})
\end{equation}
where $\rho$ is any mixed state which is a convex combination of $\{a_{i}, \ket{\psi^{i}}\}$ of pure states
$$ \rho = \sum_{i} a_{i}\ket{\psi^{i}}\bra{\ket{\psi^{i}}}$$
To find GEM of mixed state, it is important to consider the nonuniqueness of the of the pure state decomposition. Here we consider convex hull construction \cite{vollbrecht2001entanglement}, a general extension method to define GEM for mixed continuous variable states.
We briefly discuss convex hull construction since we will need this for one of main results.
Consider P be a convex set and $Q\subset P$ be an arbitrary subset. Let $F:Q \rightarrow \mathbb{R}\cup \{+\infty\}$. We then define the function $coF: P \rightarrow \mathbb{R}\cup \{+\infty\}$ by 
\begin{equation}
  coF(S) = inf\{\sum_{i}r_{i}F(x_{i})|x_{i}\to Q, \sum_{i}r_{i}x_{i} = S\}  
\end{equation}
where infimum is over all convex combinations with $\sum_{i}r_{i}\ge 0$, $\sum_{i}r_{i} =1$ and infimum over an empty set is $+\infty$.
Now consider an example, the entropy \cite{vedral1997quantifying} is, $E(a,b) = tr_{a}(lna-lb)$
In this notation, the definition of general entanglement or the relative entropy is
$$E_{RE}(\rho) = CoE (\rho)$$
We provide a general method for a class of continuous variable mixed states via above method which satisfies the following condition. An arbitrary state $\rho$ is invariant under transformation such that $\rho = \rho^{'}$.

\textbf{\textit{$\rho$ remains invariant under the transformation $\mathcal{X}\to \mathcal{\wedge}_{m}\mathcal{X}$.}} \\
Proof: one can define
\begin{equation}
    \begin{split}
        \Tilde{\rho} & = \ket{\psi}\bra{\psi^{*}} \\
         & = \int \phi (N_{1}\mathcal{X})\phi(N_{2}\mathcal{X})N_{1}\ket{\mathcal{X}}\bra{\mathcal{X}}N_{2}^{T}\textrm{d}\mathcal{X} 
    \end{split}
\end{equation}
The matrix element of $ \Tilde{\rho}$ could be written as
$$ \Tilde{\rho} = \Tilde{\Phi}(\mathcal{X})N_{1}\ket{\mathcal{X}}\bra{\mathcal{X}}N_{2}^{T}$$
Under transpose, where the transposition is done on the $\mathcal{\Tilde{M}}$ sub-system.
 $$\Tilde{\rho} {\to} \Tilde{\Phi}(\mathcal{X})N_{1}\wedge_{m}\ket{\mathcal{X}}\bra{\mathcal{X}}\wedge_{m}^{T}N_{2}^{T} = \Tilde{\rho}_{(\wedge_{m}\mathcal{X})}   $$
\begin{equation}
    \Tilde{\rho}^{'}= \int \phi (N_{1} \mathcal{X})\phi (N_{2} \mathcal{X}) N_{1} \wedge_{m} \ket{\mathcal{X}}\bra{\mathcal{X}}\wedge_{m}^{T}N_{2}^{T} d\mathcal{X}
\end{equation}
Since the integration is on $\mathcal{X}$, $\Tilde{\rho}^{'}$ does not change under the substitution $\mathcal{X} \to \wedge_{m}\mathcal{X}$.
$$\Tilde{\rho}^{'}  = \phi (N_{1}\wedge_{m}\mathcal{X})\phi (N_{2}\wedge_{m}\mathcal{X})N_{1}\ket{\mathcal{X}}\bra{\mathcal{X}}N_{2}^{T}d\mathcal{X}$$
Hence $\rho = \rho^{'}.$

In principle, one can have a set of states for which  $\rho = \rho^{'}$, then it is sufficient to perform the optimization over the set. 
If any mixed CV state satisfies the above, then this method can be successfully implemented to find the GEM for the state. Note that, this method is directly connected to other methods that we discussed in Sec. III.

Now we show that GEM is a "good" measure of entanglement \cite{vedral1997quantifying} which satisfies all the three conditions.
The following necessary conditions, the measure of entanglement $ \mathcal{G}(\rho)$ has to satisfy:\\
 1.  $ \mathcal{G}(\rho) = 0$ iff $\psi$ is separable. \\
 2. $ \mathcal{G}(\rho)$ is invariant under local unitary operations.\\
 3. The measure of entanglement cannot  increase under local general measurements (LGM) + classical communication (CC).
 
To satisfy condition 1, it is sufficient to demand that $ \mathcal{G}(\rho) = 0$, iff $\Tilde{\Phi}(
\mathcal{X}) = \Tilde{\Phi}(\wedge_{m}\mathcal{X}).$
Because of the invariance of $\rho$ under $\mathcal{X}\to \mathcal{\wedge}_{m}\mathcal{X}$, condition 2 is automatically satisfied.

\textbf{\textit{$\Tilde{\Phi}(\mathcal{X}) \Tilde{\Phi}^{*}(\wedge_{m}\mathcal{X})$ is non increasing under every completely positive, trace preserving map.}} \\
Proof: A complete measurement is given as a unitary operation + partial tracing on extended Hilbert space.
For any completely positive, trace preserving map $\sigma$ i.e., $\sigma (\mathcal{X}) = \sum W_{i}\mathcal{X}W_{i}^{\dagger}$ and $\sum_{i} W_{i}^{\dagger}W_{i} =1$

Vedral et al. \cite{vedral1997quantifying} presented a set of sufficient conditions that are written below: \\ 
(T1)  Unitary operations leave $G(\lambda || \omega )$ invariant i.e., $G(\lambda || \omega ) = G( U \lambda U^{\dagger} ||  U\omega U^{\dagger})$. \\
(T2) $G(Tr_{\omega}\lambda || Tr_{\omega}\omega ) \leq \lambda || \omega$, where $Tr_{\rho}$ is a partial trace.\\
(T3) $G(\lambda \otimes \ket{\alpha}\bra{\alpha} || \omega \otimes \ket{\alpha}\bra{\alpha} ) = G(\lambda || \omega )$. 

where $W$ is an operator satisfying the completeness relation $\sum_{i} W_{i}^{\dagger}W_{i} =1$, $G$ is any measure between two states $\lambda \: \textrm{and}\: \omega$ and defined as $G(\lambda||\omega)$.

Let's define, $V = \sum_{i}W_{i}\otimes \ket{i}\bra{\eta}$
where ${\ket{i}}$ is an orthonormal basis and $\eta$ is a unit vector.
$$V^{\dagger}V = 1 \otimes \ket{\eta}\bra{\eta}$$
There is a unitary operator U such that
\begin{equation}
    \begin{split}
       U (\mathcal{X} \otimes \ket{\eta}\bra{\eta})U^{\dagger} &= \sum_{ij} W_{i} \mathcal{X}W_{i}^{\dagger} \otimes \ket{i}\bra{j}, \\
       Tr{\{{U(\mathcal{X}\otimes \ket{\eta}\bra{\eta})U^{\dagger}\}}} &= \sum_{i}W_{i}\mathcal{X}W_{i}^{\dagger}
    \end{split}
\end{equation}

Using (T2),

\begin{align*}
    \phi \big( Tr_{2}\{U(\mathcal{X}\otimes\ket{\eta}\bra{\eta})U^{\dagger} \}\big) \phi^{*}\big( Tr_{2}\{U(\lambda_{m}\mathcal{X}\otimes\ket{\eta}\bra{\eta}U^{\dagger} \}) \\ \leq \phi ( U(\mathcal{X}\otimes\ket{\eta}\bra{\eta})U^{\dagger})  \phi^{*} (U(\lambda_{m}\mathcal{X}\otimes\ket{\eta}\bra{\eta})U^{\dagger} )
\end{align*}

Using (T3)
$$ \phi (\mathcal{X}\otimes \ket{\eta}\bra{\eta})\phi^{*} (\lambda_{m}\mathcal{X}\otimes \ket{\eta}\bra{\eta}) = \phi (\mathcal{X})\phi^{*} (\lambda_{m}\mathcal{X}) $$
This proves the above condition.\\
Our measure which has a statistical operational basis that might enable experimental determination of the quantitative degree of entanglement.
Now the GEM for multiparty CV states is straight forward.
We examine GEM for a experimentally realized state in the following section.
\subsection{Phase-matched squeezed state}
The phase randomized two-mode squeezed vacuum state \cite{kohnke2021quantum} is given by
\begin{equation}
    \rho_{2} = \sum (1-r) r^{n} \ket{n}\bra{n} \otimes \ket{n}\bra{n}
\end{equation}
where $r = \tanh{\xi}$, $\xi$, is a  complex squeezing parameter. One can observe that Eq. (29) has no entanglement because it is a convex mixture of tensor product states $\ket{n}\bra{n} \otimes \ket{n}\bra{n}$ and convex mixture is considered as classical mixture of product states.
In Eq. (29), the phase is equally distributed called fully phase randomized state.
Consider the phase is not equally distributed, the  phase-matched squeezed state is given by
\begin{equation}
    \rho^{'}_{2} = \sum_{m,n} p(m,n) (1-r)r^{\frac{m+n}{2}}\ket{m,m}\bra{n,n}
\end{equation}
where $p(m,n) = \exp [\frac{-\sigma^{2}(m-n)^{2}}{2}]$, $r = \xi^{2}$ and $\sigma$ is the variance. $\rho^{'}_{2}$ is experimentally accessible and entanglement test can be performed experimentally. The GEM is given as
\begin{equation}
    \mathcal{G}(\rho) = 2\Big[1 - \sum_{m} (1-\xi^{2})^{2}\xi^{4m}\Big]
\end{equation}

\begin{figure}[ht]
    \centering
    \includegraphics[width= 0.5\textwidth]{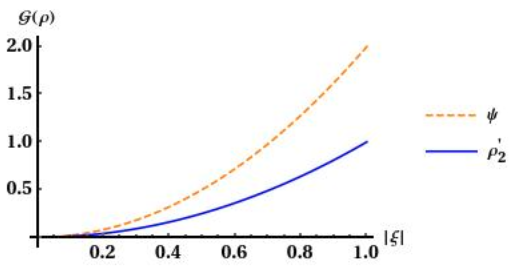}
    \caption{Variation of the entanglement measure $ \mathcal{G}(\rho)$ for phase-matched squeezed state and squeezed vacuum state with $|\xi|$, All the axes are dimensionless.}
    \label{fig:figure3}
   \end{figure}

We show the variation of the GEM $ \mathcal{G}(\rho)$ with $|\xi|$ in FIG. 3. In the case of phase-matched squeezed state, when $|\xi| = 0$, $ \mathcal{G}(\rho)$ starts increasing. Then, when $|\xi|$ increases slowly, it increases and when $|\xi|$ tends to 1, it saturates.
For better understanding of the entanglement of the phase-matched squeezed state, we have compared the GEMs of this state and of the squeezed vacuum states $\psi = \sqrt{1-|\xi|^{2}} \sum \xi^{r} \ket{r,r}$ in the same Fig. 3. Clearly,  $ \mathcal{G}(\rho)$ grows much faster and varies linearly with $|\xi|$ than that of the phase-matched squeezed state, at $|\xi| =1$, the squeezed vacuum states attains maximum value. The entanglement measure $ \mathcal{G}(\rho)$ of the squeezed vacuum state is larger than that of the phase-matched squeezed state.   



\section{Discussion and Conclusion}

One may note the dynamics in the phase space by considering the Wigner transform on both the sides of $\Tilde{\Phi}(\mathcal{X}) = \Tilde{\Phi}(\wedge_{m}\mathcal{X})$, where $\mathcal{X}_{2n \cross 1} \equiv \begin{bmatrix}
\vec{x}_{1}\\
\vec{x}_{2}
\end{bmatrix}$ with ${\vec{x}_{1}, \vec{x}_{2}}$ being some general coordinates of the system. Noting that $\vec{x}_{1}$ and $\vec{x}_{2}$ are independent, the Wigner transform of the LHS is
\begin{equation}
    \begin{aligned}
      \Tilde{W}(\mathcal{X}, \mathcal{P}) &= (\frac{1}{\pi})^{2n} \int e^{2i\mathcal{P}.\mathcal{Y}}\Tilde{\Phi}(\mathcal{X}-\mathcal{Y})\Tilde{\Phi}^{*}(\mathcal{X}+\mathcal
    {Y})d\mathcal{Y}\\
    & = W(\vec{x}_{1},\vec{p}_{1})W(\vec{x}_{2},\vec{p}_{2})
    \end{aligned}
\end{equation}
where $\mathcal{Y}\equiv \begin{bmatrix}
\vec{y}_{1}\\
\vec{y}_{2}
\end{bmatrix}$,  $\mathcal{P}\equiv \begin{bmatrix}
\vec{p}_{1}\\
\vec{p}_{2}
\end{bmatrix}$ and $W$ is the corresponding Wigner function of the given state $\ket{\psi}$. Under separability, $ \Tilde{W}(\mathcal{X}, \mathcal{P})$ may equivalently be written as
\begin{equation*}
    = (\frac{1}{\pi})^{2n} \int e^{2i\mathcal{P}.\mathcal{Y}}\Tilde{\Phi}(\wedge_{m}(\mathcal{X}-\mathcal{Y}))\Tilde{\Phi}^{*}(\wedge_{m}(\mathcal{X}+\mathcal
    {Y}))d\mathcal{Y}
\end{equation*}
and since the integration runs on $\mathcal{Y}$, transforming $\mathcal{Y} \to \wedge_{m}\mathcal{Y}$ does not change the integral. Therefore, under separability, $\Tilde{W}(\mathcal{X}, \mathcal{P})$
\begin{equation}
\begin{aligned}
  & = (\frac{1}{\pi})^{2n} \int e^{2i\mathcal{P}.\wedge_{m}\mathcal{Y}}\Tilde{\Phi}(\wedge_{m}\mathcal{X}-\mathcal{Y})\Tilde{\Phi}^{*}(\wedge_{m}\mathcal{X}+\mathcal
    {Y})d\mathcal{Y}\\
    & = \Tilde{W}(\wedge_{m}\mathcal{X},\wedge_{m}\mathcal{P}) \nonumber
    \end{aligned}
\end{equation}
noting that $\mathcal{P}.\wedge_{m}\mathcal{Y}= \wedge_{m}\mathcal{P}.\mathcal{Y}$, $d\mathcal{Y} = d(\wedge_{m}\mathcal{Y})$, and $\wedge_{m}^{2} = 1$. Therefore, $\Tilde{W}(\xi)$ being invariant under the coordinate transformation $\xi \to \begin{pmatrix}
\wedge_{m} & 0 \\
0 & \wedge_{m}
\end{pmatrix} \xi$, where $\xi_{4n\cross 1} = \begin{bmatrix}
\mathcal{X}\\
\mathcal{P}
\end{bmatrix}$, is a necessary and sufficient condition for separability.
Considering, $Tr(\rho_{PT}^{4}) = Tr[(\rho^{\mathcal{M}})^{2} \otimes (\rho^{\Tilde{\mathcal{M}}})^{2}] = (Tr[(\rho^{\mathcal{M}})^{2}])^{2}$, one may rewrite entanglement measure as
\begin{equation}
    \begin{aligned}
      \mathcal{E}_{\mathcal{M}}^{2} & = 2\Big[1 -\sqrt{Tr(\rho_{PT}^{4})}\;\Big]\\
      & = 2\Big[1-\sqrt{\int W_{PT}^{4}(\vec{x},\vec{p})d\vec{x}d\vec{p}}\;\Big]
    \end{aligned}
\end{equation}
For the case of two-degrees of freedom, therefore, if $\int W_{PT}^{4}(x_{1}, p_{1}, x_{2}, p_{2}) d\vec{x} d\vec{p} =1$ for a given pure state. One may write $W_{PT}(x_{1}, p_{1}, x_{2}, p_{2}) = W(x_{1}, p_{1}, x_{2}, -p_{2})$ as shown by Simon \cite{simon2000peres} under separability. Observing that $Tr(\rho_{PT}) =1$ and $Tr(\rho_{PT}^{2}) =1$ for any given pure density matrix $\rho$, one may note that, when $\rho_{PT}$ is positive semi-definite, the eigen values must be either 0 or 1 with multiplicity one in the DV case. So any higher powers of $\rho_{PT}$ would also have unit trace. Hence, $\rho$ is separable iff $\rho_{PT}$ is positive-semi-definite.\\

We conclude by commenting on the connections to other widely used measures in the literature to show their equivalence to the GEM. The Hilbert-Schmidt distance D between two density matrices $\rho_{1} \: \textrm{and} \: \rho_{2}$ is $D_{\rho_{1}}^{2}(\rho_{2}) \equiv ||\rho_{1} - \rho_{2}||^{2}_{HS} = Tr[(\rho_{1} -\rho_{2})^{2}]$ has been widely used to study the geometry and structure of entanglement with connections to negativity and PPT-states \cite{verstraete2002geometry, banerjee2019minimum}.
Now using Eq.(3) and (4)\\

\begin{widetext}
\begin{equation}
  \Tilde{\rho} - \Tilde{\rho_{PT}} = \int \big[ \phi(\mathcal{N}_{1}\mathcal{X})\phi(\mathcal{N}_{2}\mathcal{X}) - \phi(\mathcal{N}_{1}\wedge_{m}\mathcal{X})\phi(\mathcal{N}_{2}\wedge_{m}\mathcal{X})\big]\mathcal{N}_{1}\ket{\mathcal{X}}\bra{\mathcal{X}}\mathcal{N}_{2}^{T}\, d\mathcal{X}   
\end{equation}
\end{widetext}
It is evident that the GEM can be instead interpreted as
\begin{equation}
    \mathcal{E}_{\mathcal{M}}^{2} =  ||\Tilde{\rho_{1}} - \Tilde{\rho_{2}}||^{2}_{HS}
\end{equation}
This would be shown below to be related to the Hilbert-Schmidt distance of the reduced density matrix from the maximally mixed state using Lagrange's identity.
Consider the case of general DV system with density matrix $\rho_{N\cross N}$. 
Taking $\rho_{1} = \frac{1}{N}1_{N\cross N}$, one may write the distance of $\rho$ to the maximally mixed state $\rho_{1}$ as
$$D^{2}(\rho) = Tr(\frac{1}{N^{2}}1 + \rho^{2} -\frac{2}{N}\rho) = Tr(\rho^{2})- \frac{1}{N}$$
noting $Tr(\rho) = 1$ and $ Tr(1) =N$. In the CV case as $N \to \infty,\, D^{2}(\rho)  = Tr(\rho^{2}) = 1- E^{2}(\rho)/2$, where E is the generalized entanglement measure. One may conversely use this property to geometrically define define a maximally mixed CV states, noting from Eq. (13), one can has the following identity for CV states
$$||\Tilde{\rho_{1}} - \Tilde{\rho_{2}}||^{2}_{HS} + 2||\rho_{\mathcal{M}}- \rho_{1}||^{2}_{HS} =2 $$
On the same note, one can show the equivalence of the von Neumann entropy as an entanglement measure to the GEM. The entropy of a density matrix $\rho$ is defined as $S = -Tr(\rho\,\textrm{ln} \, \rho) = \langle \textrm{ln}\, \rho \rangle$, where $\langle . \rangle$ denotes the expectation value. Expanding S around a pure state $\rho^{2} = \rho$, that is, the non-negative matrix $1 - \rho$, and noting that $E^{2}/2 = 1 - Tr(\rho^{2}) = Tr(\rho(1-\rho)) = \langle 1-\rho \rangle$, one infers
$$S = -\langle \textrm{ln}\, \rho \rangle = \langle 1-\rho \rangle + \langle (1-\rho)^{2}/2 \rangle + \langle (1-\rho)^{3}/3 \rangle + ...,$$
and therefore, $ S = \frac{E^{2}}{2} + $ residual. Clearly, iff E = 0 , the residual term vanish, giving S = 0; else when $E > 0$, the residual remains positive, giving $S > E^{2}/2$ for any state $\rho$. Therefore. S and E are equivalent in characterizing separable states and entanglement among entangled states faithfully. It is, however, faster computationally to calculate the GEM than it is to find the von Neumann entropy, as it does not require diagonalization of the density matrix.

The convex roof construction involves optimization and is usually hard. 
The entanglement of an arbitrary mixed continuous variable state is not a simple task. In this paper, we defined entanglement for pure continuous variable states and extending the concept to mixed continuous variable states via the convex roof construction. We evaluated the measure for several class of continuous variable states. However, it is not clear whether the same method is useful for the mixture of states which have white or colored noise. The persistence of sub-planck structure in a mixed continuous variable states is being possible with a specific environmental conditions \cite{kumari2015sub}.  
We hope our work provides new insights into the geometry and structure of entanglement in general both in pure and mixed continuous variable systems by proving a family of faithful entanglement measures, and equivalent forms of necessary and sufficient conditions for separability across arbitrary bipartitions. We believe the results hold deep connections to the recent works on the nature of quantum correlations in many-body systems \cite{reiter2016scalable}, monogamy of entanglement \cite{coffman2000distributed, allen2017polynomial}, and fundamental aspects of quantum mechanics, including, the uncertainty principle and commutation relations \cite{simon2000peres, werner2001bound,duan2000inseparability}. 

\textbf{Acknowledgements}:
SNS and VSB equally contributed to this work.

SNS is thankful to the University Grants Commission and Council of Scientific and Industrial Research, New Delhi, Government of India for Junior Research Fellowship at IISER Kolkata. VSB contributed to this article in his personal capacity, and the conclusions reached are his own and do not represent the views of Samsung Research America, Inc. PKP acknowledges the support from DST, India through Grant No. DST/ICPS/QuST/Theme-1/2019/2020-21/01.

\appendix
\section{Proof of Lagrange's identity}
Considering RHS of Eq. (11),
 \begin{equation*}
     \begin{aligned}
       & = \iint_{\vec{t}}^{\vec{u}} \Big|f(\vec{x}) g(\vec{x^{'}}) - f(\vec{x^{'}}) g(\vec{x})\Big|^{2} d\vec{x^{'}} d\vec{x}  \\
       & = \frac{1}{2} \iint_{\vec{t}}^{\vec{u}} \Big|f(\vec{x}) g(\vec{x^{'}}) - f(\vec{x^{'}}) g(\vec{x})\Big|^{2} d\vec{x^{'}} d\vec{x} \\
       & = \frac{1}{2}\iint_{\vec{t}}^{\vec{u}} \big[f(\vec{x}) g(\vec{x^{'}}) - f(\vec{x^{'}}) g(\vec{x})\big] \\ &\big[f^{*}(\vec{x}) g^{*}(\vec{x^{'}}) - f^{*}(\vec{x^{'}}) g^{*}(\vec{x})\big]  d\vec{x^{'}} d\vec{x} 
     \end{aligned}
 \end{equation*}
\begin{multline*}
    = \frac{1}{2} \iint_{\vec{t}}^{\vec{u}} \Big[ |f(\vec{x})|^{2} |g(\vec{x^{'}})|^{2} -2 \textrm{Re}(f(\vec{x}) g(\vec{x^{'}})f^{*}(\vec{x^{'}}) g^{*}(\vec{x})) \\+ |f(\vec{x^{'}})|^{2} |g(\vec{x})|^{2}\Big] d\vec{x^{'}} d\vec{x} 
\end{multline*}
\begin{multline*}
    = \Big(\int_{\vec{t}}^{\vec{u}} |f(\vec{x})|^{2} d\vec{x}\Big)\Big(\int_{\vec{t}}^{\vec{u}} |g(\vec{x^{'}})|^{2}d\vec{x}\Big) - \\ \textrm{Re}\iint_{\vec{t}}^{\vec{u}} f(\vec{x}) g(\vec{x^{'}})f^{*}(\vec{x^{'}}) g^{*}(\vec{x}) d\vec{x^{'}} d\vec{x}
\end{multline*}
\begin{multline*}
     = \Big(\int_{\vec{t}}^{\vec{u}} |f(\vec{x})|^{2} d\vec{x}\Big)\Big(\int_{\vec{t}}^{\vec{u}} |g(\vec{x^{'}})|^{2}  d\vec{x}\Big) - \Big|\int_{\vec{t}}^{\vec{u}} f(\vec{x}) g^{*}(\vec{x}) d\vec{x}\Big|^{2} = \textrm{LHS}
\end{multline*}
Hence the identity.

\bibliography{sns.bib}

\end{document}